\documentclass[twocolumn,aps,showpacs,amsmath,amssymb]{revtex4}

\usepackage{graphicx}
\usepackage{dcolumn}
\usepackage{bm}

\begin{document}

\title{Nuclear reactions in hot stellar matter and nuclear
surface deformation}

\author{V. Yu. Denisov}
\address{%
Institute for Nuclear Research, Prospect Nauki 47,
03680 Kiev, Ukraine }%

\date{\today}

\begin{abstract}
Cross-sections for capture reactions of charged particles in hot stellar matter turn out be increased by the quadrupole surface oscillations, if the corresponding phonon energies are of the order of the star temperature. The increase is studied in a model that combines barrier distribution induced by surface oscillations and tunneling. The capture of charged particles by nuclei with well-deformed ground-state is enhanced in stellar matter. It is found that the influence of quadrupole surface deformation on the nuclear reactions in stars grows, when mass and proton numbers in colliding nuclei increase.
\end{abstract}


\maketitle

\section{Introduction}

Various nuclear reactions take place in the stars \cite{st,rr,bethe,lw,whw,ns1,rp,ra,ns3,ns2,sr,s,r,oo,fragm,beta}. The cross-sections of nuclear reactions determine diverse properties of the stars \cite{st,rr,bethe,whw,oo} and the nucleosynthesis of elements in stellar matter \cite{st,rr,bethe,lw,whw,ns1,rp,ra,ns3,ns2,sr,s,r}. 
Nuclear reactions occur at very high temperatures during the burning of massive stars, supernova explosions or in the crust of neutron stars \cite{st,bethe,lw,whw}. A typical temperature of silicon burning of massive stars is close to $3.5 \cdot 10^9$ K $\approx 0.3$ MeV \cite{whw}. The star matter at temperatures $\sim 0.3$ MeV mainly consists of the most tightly bound iron-group nuclei, $\alpha$-particles, nucleons, electrons and $\gamma$-quanta \cite{st,rr,lw,whw,ns2}. Subsequent stages of the evolution of massive stars may take place at higher temperatures. The composition, density and temperature of stellar matter depend on the mass, as well as on the type and age of the star \cite{st,rr,bethe,lw,whw}.

Nuclei in the star participate in various reactions induced by $\gamma$-quanta, electrons, free nucleons and other nuclei. Therefore nuclei in star matter exist in both the ground and excited states due to photoexcitation, inelastic collisions between the nuclei and electrons, free nucleons and other nuclei. At star temperatures $\gtrsim$ 0.2 MeV the population probability of low-energy ($0.2 \div 1$ MeV) excited states in soft nuclei is rather noticeable. Such states are exemplified by the lowest $2^+$ surface oscillation states in  $^{52}$Fe and  $^{80}$Sr with excitation energies, respectively, of $\varepsilon_{\rm vib}=0.849$ MeV and 0.386 MeV \cite{be2}. The total vibrational amplitudes of these states are large, namely, $\beta_{\rm vib}=0.308$ and $0.404$ \cite{be2}. Therefore evaluating the reaction cross-sections between charged particle and heavy soft nucleus in star matter we should take into account contributions from both the ground and well-deformed excited states of the nucleus. 

The astrophysical importance of charged-particle capture on target nuclei with $N \cong Z$ is manifold. These capture reactions are important for nucleosynthesis processes in stellar matter and various burning phases of massive stars \cite{rr,lw,whw,rp,ra}.

Cross-sections of nuclear reactions may be evaluated using the nucleus-nucleus or nucleon-nucleus potentials \cite{rr,satchler,fl}. The nucleus-nucleus potential depends on the shape of nuclei participating in the reaction \cite{satchler,fl,dn,wong,dr,di,ccdef}. In the case of interaction between spherical and prolate nuclei, the barrier for the tip orientation of deformed nucleus is smaller than one for the side orientation. For example, these barriers for system $^{48}$Ca+$^{238}$U are respectively close to 184 MeV and 202 MeV for corresponding orientation of $^{238}$U \cite{dn}, while the value of the barrier evaluated for spherical shape of $^{238}$U is 197 MeV \cite{d}. Similar reduction of the barrier induced by deformation strongly enhances the sub-barrier fusion cross-sections for light and medium-weight collision systems \cite{fl,wong,dr,di,ccdef}. Note that charged-particle capture on heavy nuclei in stellar matter takes place at sub-barrier energies at the temperatures $\lesssim0.5$ MeV. Therefore the enhancement of capture rates  caused by deformation of heavy nuclei should be important for various reactions in stars.

The $2^+$ shape oscillations in soft nuclei may also affect nuclear reaction rates in the hot stars, because the energies of such states are small and low-energy nuclear states can be appreciably populated in stellar matter at temperatures $\gtrsim 0.2$ MeV. The amplitude of surface deformation in soft nuclei is large. Consequently, the reduction of fusion barrier induced by this deformation can be significant. Cross-sections of fusion (capture) reaction in stars can be enhanced essentially by the barrier reduction caused by $2^+$ shape oscillations. Therefore quadrupole shape oscillations in soft nuclei should be taken into account for accurate evaluation of reaction cross-section in hot star matter. 

Influence of star environment and specific properties of nuclei on nuclear reactions in stars have been taken into account in some calculations in the nuclear astrophysics \cite{ns1,s,r}. However the $p$- and $\alpha$-capture reactions on deformed nuclei are usually treated by means of an effective spherical potential of equal volume \cite{rp,ra,r}. We consider enhancement of cross-sections for various nuclear reactions in hot stellar matter induced by the low-energy $2^+$ oscillations of nuclear shape, because the barrier between nuclei is reduced due to shape deformation induced by the $2^+$ oscillations. As the result the probability of sub-barrier tunneling is strongly increased at smaller values of the barrier height. Such mechanism of the nuclear reactions enhancement in stars has not been considered yet. Our consideration includes $\alpha$- and $p$-capture, as well as heavy-ion fusion reactions for soft nuclei.

Reactions $\alpha$+Ne, $\alpha$+Mg, $\alpha$+Si and some other are very important for the O-Si burning phases of massive stars \cite{st,rr,lw,whw,ra}. Heavy nuclei participating in these reactions are well-deformed in the ground-state. Many other proton-rich nuclei participating in various charged-particle capture reactions in stellar matter are well deformed in the ground-state. We include the ground-state deformation of heavy nucleus into the interaction potentials between charged particle and heavy nucleus. In the framework of such treatment we discuss cross-section enhancement caused by the ground-state deformation of heavy nucleus for various reactions in stellar matter.

This article is organized as follows. Partition probabilities of the ground and $2^+$ surface oscillation states of nuclei in stellar matter are discussed in Sec. II. Influence of low-energy quadrupole vibrations of nuclear surface on both  $S$-factors and the velocity-averaged cross-sections of $\alpha$-nucleus capture reactions in stars is considered in Sec. III. The effect of $2^+$ surface oscillations on proton capture and nucleus-nucleus fusion reactions in stellar matter is studied in Sec. IV and V, respectively. The capture of charged particles by nuclei with well-deformed ground-state is discussed in Sec. VI. Conclusions are presented in Sec. VII.

\section{Partition probability of nuclear states in stellar matter}

The probability to find a nucleus in a state with excitation energy $\varepsilon_i$ and spin $j_i$ in stellar matter at temperature $T$ can be estimated within the statistical approach as
\begin{eqnarray}
	P(\varepsilon_i,j_i,kT)= \frac{(2j_i+1) \exp(-\varepsilon_i/kT)}{\sum_{i=0}^\infty (2j_i+1) \exp(-\varepsilon_i/kT)} . 
\end{eqnarray}
Here $k$ is the Boltzmann constant. We use $i=0$ for the ground-state of the nucleus with $\varepsilon_i=0$, $i=1$ for the lowest $2^+$ surface oscillation state with $\varepsilon_1=\varepsilon_{\rm vib}$ and $i \geq 2$ for other excited states. 

\begin{figure}
\includegraphics[width=9cm]{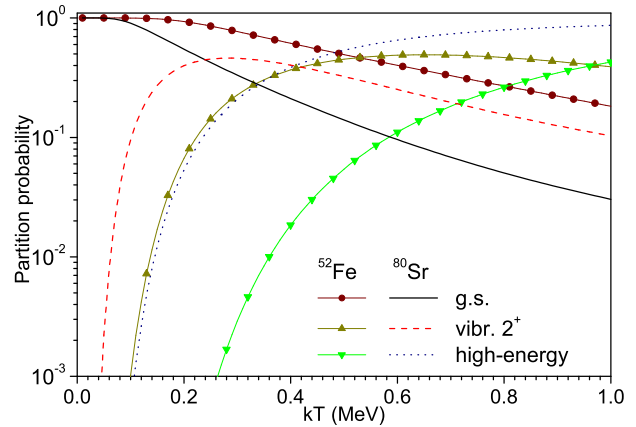}

\caption{(Color online) Occupation probability for the ground state, the first $2^+$ surface oscillation state and the net occupation probability of high-energy states $P(\varepsilon_i>\varepsilon_{\rm vib})$ in  $^{52}$Fe and  $^{80}$Sr at different temperatures of stellar matter $kT$.}
\end{figure}

The partition probabilities of both the ground and the lowest $2^+$ surface oscillation states in  $^{52}$Fe and  $^{80}$Sr nuclei for different temperatures of stellar matter are presented in Fig. 1. The partition probabilities are evaluated by using the experimental data for the lowest 98 states in  $^{52}$Fe (up to 11.780 MeV) and 90 states in  $^{80}$Sr (up to 15.576 MeV) listed in compilation \cite{nudat}. In addition to energies, values of spin for many levels are also given in this data compilation \cite{nudat}. If the value of spin for level $i$ is not listed then we assign $j_i=0$. On the other hand, when the spin is not well defined from experiments and several possible values of spin are listed for a level in Ref. \cite{nudat}, we choose the maximal value among the listed values. It should be mentioned that this choice for spin assignments is of marginal practical importance for specific population probabilities displayed in Fig. 1. Those depend primary on the spins of the lowest states. The spins of 6 and 3 lowest states in $^{52}$Fe and  $^{80}$Sr, respectively, are well established \cite{nudat}.

The total occupation probability of high-energy states with energies $\varepsilon_i>\varepsilon_{\rm vib}$ is equal to 
$$P(\varepsilon_i>\varepsilon_{\rm vib})=1-\sum_{i=0,1} P(\varepsilon_i,j_i,kT).$$ We see in Fig. 1 that the net occupation probability of high-energy states becomes dominant at very high temperatures of the stellar matter, $kT \gtrsim 0.5$ MeV. As seen further in Fig. 1 the partition of $2^+$ surface oscillation states is negligible for small temperatures $kT \lesssim 0.1$ MeV. On the other hand, the $2^+$ vibrational states are noticeably populated at temperatures $0.2$ MeV $\lesssim kT \lesssim 0.7$ MeV. Moreover, the occupation for vibrational states is higher than for the ground states at temperatures $kT > 0.53$ MeV for  $^{52}$Fe and at $kT > 0.24$ MeV for  $^{80}$Sr. Therefore, nuclear reaction cross-sections at moderate temperatures of stellar matter should be evaluated by taking into account properties of both the ground and $2^+$ states. Nuclei  $^{52}$Fe and  $^{80}$Sr are spherical in the ground state but generally deformed in the first $2^+$ surface vibrational states. Thus it is necessary to consider the nuclear reactions both for spherical and deformed shapes of nuclei  $^{52}$Fe and  $^{80}$Sr in the hot star matter.

The cases of the nuclei  $^{52}$Fe and  $^{80}$Sr are interesting from the point of view of the role of deformation in reactions. On the one hand, energies of the $2^+$ surface oscillations $\varepsilon_{\rm vib} \equiv \varepsilon_1$ are low, making the occupations for these states significant in the hot stellar medium. On the other hand, the stiffness of nuclear surface with respect to  deformations $C_{\rm vib}$ are also low for these nuclei, making the shape vibration amplitudes $\beta_{\rm vib}=\left(5 \varepsilon_{\rm vib}/(2C_{\rm vib})\right)^{1/2}$ large. We consider $2^+$ surface oscillations in nuclei in the framework of the harmonic oscillator model \cite{soloviev}.

There is a variety of nuclei in the stellar matter. Nuclei far form the $\beta$-stability line are soft as a rule. Nuclei with number of nucleons, being far from the magic numbers, and lying along $\beta$-stability line are often soft too. However, there are also many rigid nuclei, such as $^{4}$He, $^{12}$C, $^{16}$O, $^{40,48}$Ca, $^{208}$Pb, appearing in the hot stellar matter. Energies of the $2^+$ surface oscillation states are rather high in those nuclei, making the $2^+$ states weakly populated in the stellar matter. Shape vibration amplitudes are small in rigid nuclei. As a result, the influence of the $2^+$ surface vibrations for such rigid nuclei on nuclear reactions in stars may be neglected. Overall, it is necessary to develop a formalism for nuclear reactions in the stellar matter, which is applicable to both to soft and rigid nuclei. 

\section{$\alpha$-capture reactions in stars}

The $\alpha$-particle may be considered the most rigid nucleus, because the energy of the first excited state is 20.21 MeV \cite{nudat}. The population of such high-energy state is negligible in stellar matter at temperatures $kT \lesssim 1$ MeV. On the other hand, many states with energies $\varepsilon_i \lesssim 1$ MeV in soft nuclei can be noticeably populated at $kT \lesssim 1$ MeV. Therefore, evaluating $\alpha$-capture reaction cross-section in stellar matter at $kT \sim 0.3$ MeV, we should take into account the contributions of various populated states in soft nuclei only. Correspondingly, the $\alpha$-capture reaction cross-section in star matter at temperature $T$ can be estimated as
\begin{eqnarray}
\sigma(E,kT)= \sum_{i=0}^{\infty} P(\varepsilon_i,j_i,kT) \sigma_i(E),
\end{eqnarray}
where $\sigma_i(E)$ is the fusion cross-section between the $\alpha$-particle and a nucleus in a state $i$ with energy $\varepsilon_i$ and spin $j_i$, $E$ is the collision energy.

We shall consider the $\alpha$-capture on nuclei  $^{52}$Fe and  $^{80}$Sr as example. Reaction $\alpha$+$^{52}$Fe$\Rightarrow ^{56}$Ni may take place in the last stage of the silicon burning of the stellar matter, when iron nuclei are the most abundant nuclei \cite{whw,ns1}. Reaction $\alpha$+$^{80}$Sr$\Rightarrow ^{84}$Zr is important for the nucleosynthesis. Note that $\alpha$-capture reactions on nuclei with $N \cong Z$ are substantial for nucleosynthesis processes in stellar and explosive burning \cite{whw,ns1,ra,rp}. 

As we have seen in the previous section, the ground state and the first $2^+$ surface oscillation state of a heavy nucleus are mainly populated in star matter at moderate temperatures. The nuclei $^{52}$Fe and $^{80}$Sr are spherical in the ground-state and generally deformed in the first $2^+$ state. 

The $\alpha$-capture reaction cross-section in star matter at temperature $T$ consists of three terms 
\begin{eqnarray}
\sigma(E,kT)= P(0,0,kT) \sigma_0(E) + P(\varepsilon_{\rm vib},2,kT)\sigma_1(E) \nonumber \\ + \sum_{i=2}^{\infty} P(\varepsilon_i,j_i,kT) \sigma_i(E), 
\end{eqnarray}
where $P(0,0,kT)$ and $P(\varepsilon_{\rm vib},2,kT)$ are the partition probabilities for the ground state and the lowest $2^+$ surface-oscillation state, respectively, while $\sigma_0(E)$ and $\sigma_1(E)$ are the cross-sections of $\alpha$-capture on heavy nucleus in the ground state and the lowest $2^+$ vibrational state correspondingly.

The fusion cross-section of two particles with corresponding values of spins 0 and $j$ is given by \cite{satchler}
\begin{eqnarray}
\sigma(E)=\frac{\pi \hbar^2}{2\mu E(2j+1)} \sum_{J\ell\ell'} 
(2J+1) t_{J\ell\ell'}(E),
\end{eqnarray}
where $\mu$ is the reduced mass, $\ell$ and $\ell'$ are the orbital moment of ingoing and outgoing channels, $J$ is the total angular momentum and $t_{J\ell\ell'}(E)$ is the generalized transmission coefficient. If there is no spin-orbit forces in the potential, then the transmission coefficient becomes independent on $J$ and $t_{J\ell\ell'}(E) = \delta_{\ell\ell'} \, t(E,\ell)$ \cite{satchler}. In this case Eq. (4) can be simplified 
\begin{eqnarray}
\sigma(E)&=&\frac{\pi \hbar^2}{2\mu E(2j+1)} \sum_{J\ell} 
(2J+1) t(E,\ell) \nonumber \\ &=& \frac{\pi \hbar^2}{2\mu E(2j+1)} \sum_{\ell} 
 t(E,\ell) \sum_{J=|\ell-j|}^{\ell+j} {(2J+1)} \nonumber \\
&=& \frac{\pi \hbar^2}{2\mu E} \sum_{\ell} 
(2\ell+1) t(E,\ell) ,
\end{eqnarray}
where $t(E,\ell)$ is the transmission coefficient for particles penetration through the interaction potential barrier between them at collision energy $E$ and orbital momentum $\ell$.

Any indications on the spin-orbit interaction between $\alpha$-particle and heavy nucleus are not know yet. The ground-state spin of $\alpha$-particle is zero. Therefore we apply Eq. (5) for evaluation of the $\alpha$-capture cross-section on heavy nucleus in the ground and excited states.

The ground-state shape of $\alpha$-particle,  $^{52}$Fe and  $^{80}$Sr is spherical, therefore the $\alpha$-capture cross-section can be evaluated in the usual manner (see Eq. (5) and Refs. \cite{satchler,fl,di,subfus}) 
\begin{eqnarray}
\sigma_0(E)=\frac{\pi \hbar^2}{2\mu E} \sum_\ell
(2\ell+1) t_0(E,\ell).
\end{eqnarray}
Here $t_0(E,\ell)$ is the transmission coefficient for penetration through the interaction potential barrier between spherical nucleus in the ground state and $\alpha$-particle at collision energy $E$. 

The cross-section of $\alpha$-capture on heavy nucleus in the $2^+$ vibrational state is also obtained by using Eq. (5). However the surface of the nucleus in $2^+$ state oscillates around spherical equilibrium shape. Therefore we should take into account both the orientation and the surface oscillation effects at evaluation of the transmission coefficient in this case.

During the $\alpha$-nucleus fusion reaction the $\alpha$-particle can arrive from any direction, therefore we should make averaging over space angles. In the case of collision between spherical and axially-deformed nuclei the averaging over space angle is reduced to the averaging over the angle $\theta$, where $\theta$ is the angle between the symmetry axis of axially-symmetric deformed nuclei and the vector directed from the center of the deformed nucleus to the center of the $\alpha$-particle. Therefore the fusion reaction cross-section between the nucleus with axial quadrupole surface vibrations and $\alpha$-particle at collision energy $E$ equals to
\begin{eqnarray}
\sigma_1(E)=\frac{\pi \hbar^2}{2\mu E} \sum_\ell
(2\ell+1) \int_0^{\pi/2} t_1(E,\ell,\theta) \sin(\theta) d\theta
\end{eqnarray}
(see also Ref. \cite{di}). Here $t_1(E,\ell,\theta)$ is the transmission coefficient, which shows the probability of penetration through the potential barrier for $\alpha$-particle incoming at angle $\theta$. 

Nuclear surface in $2^+$ state in  $^{52}$Fe and  $^{80}$Sr oscillates about the spherical equilibrium shape. The distance between the deformed nuclear surface and the origin is
\begin{eqnarray}
	R(\theta)=R_0(1 + \beta Y_{20}(\theta)),
\end{eqnarray}
where $R_0$ is the radius of spherical nucleus, $\beta$ is the deformation parameter and $Y_{20}(\theta)$ is the spherical harmonic function. The distribution of deformation parameter values in $2^+$ state is described by square of the vibrational wave function $\varphi_{\rm vib}(\beta)$
\begin{eqnarray}
D(\beta) = |\varphi_{\rm vib}(\beta)|^2 = \frac{\beta^2}{\sqrt{2\pi} \beta_{0}^3} \exp{\left(-\frac{\beta^2}{2\beta_{0}^2}\right)},
\end{eqnarray}
where $\beta_{0}=\beta_{\rm vib}/\sqrt{5}$ is the zero-point amplitude. Here we use the harmonic oscillator model of the $2^+$ surface vibrational state in nuclei \cite{soloviev}. The value of $\beta$ can be different during the barrier penetration, therefore we should make averaging over all possible values of the deformation parameter. As a result
\begin{eqnarray}
t_1(E,\ell,\theta)=\int_{-\infty}^{\infty} D(\beta) \, t_1(E,\ell,\theta,\beta) \; d \beta,
\end{eqnarray}
where $t_1(E,\ell,\theta,\beta)$ is the transmission coefficient of a charged particle incoming at angle $\theta$ and evaluated at the value of surface deformation $\beta$. (Here we consider the evaluation of cross-section and other related quantities in the framework of the time-independent scattering theory \cite{fl}. Therefore the time-independent oscillator wave function $\varphi_{\rm vib}(\beta)$ is used in Eqs. (9)-(10). The averaging over  $\beta$ in Eq. (10) is equivalent to the time averaging over the period of the nuclear surface oscillations in the framework of time-dependent scattering theory \cite{fl}.)

We estimate the transmission coefficients in Eqs. (6) and (10) using the semi-classical WKB approximation at collision energies below barrier
\begin{eqnarray}
t_0(E,\ell) = \{1
\;\;\;\;\;\;\;\;\;\;\;\;\;\;\;\;\;\;\;\;\;\;\;\;\;\;\;\;\;\;
\;\;\;\;\;\;\;\;\;\;\;\;\;\;\;\;\;\;\;\;\;\;\;\;\;\;\;\;\;\;
\\ + \left. \exp\left[\frac{2}{\hbar}
\int_{a}^{b} dr \sqrt{2\mu
\left(v_0(r,\ell,E)-E\right)} \right]\right\}^{-1}, \nonumber\\
t_1(E,\ell,\theta,\beta) = \{1
\;\;\;\;\;\;\;\;\;\;\;\;\;\;\;\;\;\;\;\;\;\;\;\;\;\;\;\;\;\;
\;\;\;\;\;\;\;\;\;\;\;\;\;\;\;\;\;\;\;\;\;\;\;\;\;\;\;\;\;\;
\\ + \left. \exp\left[\frac{2}{\hbar}
\int_{a(\theta)}^{b(\theta)} dr \sqrt{2\mu
\left(v_1(r,\ell,\theta,E,\beta)-E\right)} \right]\right\}^{-1}. \nonumber
\end{eqnarray}
Here $v_0(r,\ell,E)$ and $v_1(r,\ell,\theta,E,\beta)$ are the interaction potentials between $\alpha$-particle and nucleus in the spherical ground-state and deformed $2^+$ state respectively, $r$ is the distance between the mass centers of colliding particles, $a, a(\theta)$ and $b,b(\theta)$ are the inner and outer turning points determined from the corresponding equations $v_0(r,\ell,E)|_{r=a,b}=E$ and $v_1(r,\ell,\theta,E,\beta)|_{r=a(\theta),b(\theta)}=E$. The transmission coefficients $t_0(E,\ell)$ and $t_1(E,\ell,\theta,\beta)$ are approximated by an expression for a parabolic barrier at collision energies higher then the barrier energy. 

We propose that parameters of $\alpha$-nucleus interactions are the same for the ground and vibrational states, therefore $v_0(r,\ell,E)=v_1(r,\ell,\theta,E,\beta=0)$. Due to this we omit indexes $0$ and $1$ for potentials below. 

The interaction potential between deformed nucleus and charged particle $v(r,\ell,\theta,\beta)$ consists of Coulomb $v_C(r,\theta,\beta)$, nuclear $v_N(r,\theta,\beta)$ and centrifugal $v_\ell(r)$ parts, 
\begin{eqnarray}
v(r,\ell,\theta,E,\beta)=v_C(r,\theta,\beta) + v_N(r,\theta,E,\beta) + v_\ell(r) . \;
\end{eqnarray}

The Coulomb part of interaction potential between $\alpha$-particle and deformed nucleus takes into account the effect of deformation to the first order 
\begin{eqnarray}
v_C(r,\theta,\beta) = \frac{z Z e^2}{r} \left[1 + \frac{3R_0^2}{5r^2}
\beta Y_{20}(\theta) \right],
\end{eqnarray}
if $r \ge r_m(\theta,\beta)$, and
\begin{eqnarray}
v_C(r,\theta,\beta) &\approx& \frac{z Z e^2}{r_m(\theta,\beta)}
\left[\frac{3}{2}-\frac{r^2}{2r_m(\theta,\beta)^2} \right. \\ & + & \left.
\frac{3R_0^2}{5r_m(\theta,\beta)^2} \beta Y_{20}(\theta) \left(2-\frac{r^3}{r_m(\theta,\beta)^3}
\right) \right], \nonumber
\end{eqnarray}
if $r\lesssim r_m(\theta,\beta)$. Here $z=2$ is the charge of $\alpha$-particle, $Z$ is the number of protons in the nucleus and $r_m(\theta,\beta)$ is the effective radius of the nuclear part of $\alpha$-nucleus potential. The inner turning point $a(\theta)$ is close to $r_m(\theta,\beta)$, therefore the representation of Coulomb field in the form (15) at distances $r \lesssim r_m(\theta)$ ensures continuity of the Coulomb field and its derivative at $r=r_m(\theta,\beta)$ \cite{di}.

The nuclear part of $\alpha$-nucleus interaction is taken in the Woods-Saxon shape 
\begin{eqnarray}
v_N(r,\theta,E,\beta) = \frac{V(A,Z,E)}{ 1
+\exp[(r-r_m(\theta,\beta))/d] }
\end{eqnarray}
with parameters \cite{di}
\begin{eqnarray}
V(A,Z,E)=-(30.275 - 0.45838 Z/A^{1/3} \\
+ 58.270 I - 0.24244 E ), \nonumber \\
r_m(\theta,\beta)=1.5268 + R(\theta) =1.5268 \\ + R_0(1 + \beta Y_{20}(\theta)) , \nonumber \\
R_0=R_p (1+3.0909/R_p^2) + 0.12430 t,\\
R_p=1.24 A^{1/3} (1 + 1.646 /A - 0.191 I), \\
t=I-0.4 A/(A+200), \\
d=0.49290 .
\end{eqnarray}
Here $A$ is the number of nucleons in a nucleus and $I=(A-2Z)/A$. 

The rotational part of the interaction is 
\begin{eqnarray}
v_\ell(r)= \hbar^2 \ell (\ell+1)/(2\mu r^2).
\end{eqnarray}

Data for both the $\alpha$-decay half-lives and the fusion cross-sections around the barrier for reactions $\alpha$+$^{40}$Ca, $\alpha$+$^{59}$Co, $\alpha$+$^{208}$Pb are well described by using this parametrization of the $\alpha$-nucleus potential \cite{di}. 

The values of the potential barrier $v_{\rm bar}(\theta)$ between spherical and deformed $(\beta \neq 0)$ nuclei at various angles $\theta$ and $\ell=0$ obey inequalities 
\begin{eqnarray}
 v_{\rm bar}(\theta=0)|_{\beta>0} < v_{\rm bar}^{\rm sph} < v_{\rm bar}(\theta=\pi/2)|_{\beta>0} , \\
 v_{\rm bar}(\theta=0)|_{\beta<0} > v_{\rm bar}^{\rm sph} > v_{\rm bar}(\theta=\pi/2)|_{\beta<0} ,
\end{eqnarray}
where $v_{\rm bar}^{\rm sph}$ is the barrier of the interaction potential if both nuclei are spherical. The barrier of the interaction potential $v(r,\ell,\theta,E,\beta)$ is reduced by quadrupole surface distortion in one or both interacting nuclei. As an example, the values of the barrier for reactions $\alpha+^{52}$Fe and $\alpha+^{80}$Sr for the spherical ground state at 15 MeV collision energy are $v_{\rm bar}^{\rm sph}=7.98$ and 11.13 MeV, respectively. If nuclei  $^{52}$Fe and  $^{80}$Sr are deformed (let $\beta=\beta_{\rm vib}$), these barrier values are distributed over the ranges $v_{\rm bar}(\theta=0,\beta=0.308)=7.35$ MeV $\div \; v_{\rm bar}(\theta=\pi/2,\beta=0.308)=8.30$ MeV and $v_{\rm bar}(\theta=0,\beta=0.404)=9.97$ MeV $ \div \; v_{\rm bar}(\theta=\pi/2,\beta=0.404)=11.70$ MeV, correspondingly. The reduction of the barrier height by $\sim 0.5 \div 1$ MeV induced by deformation $\beta=\beta_{\rm vib}$ increases the transmission coefficient at sub-barrier energies for $\theta$ close to 0. As a result, the fusion cross-section is strongly enhanced for sub-barrier collision energies $E$. 

The shape of a nucleus in highly-excited states $i \geq 2$ can be spherical or deformed. The kind of surface deformation can be different for different high-energy states. The high-multipolarity $\lambda \geq 3$ axial or nonaxial multipole $\lambda \geq 2$ nuclear surface deformations usually lead to the smaller reduction of the barrier than those induced by the axial quadrupole surface deformation. Therefore we can approximate $\sigma_i(E)|_{i \geq 2} \approx \sigma_0(E)$ and apply our model to high stellar temperatures $kT \lesssim 1$ MeV. As a result, the $\alpha$-capture reaction cross-section in star matter (3) can be rewritten as
\begin{eqnarray}
\sigma(E,kT) \approx [P(0,0,kT)+\sum_{i=2}^{\infty} P(\varepsilon_i,j_i,kT)] \sigma_0(E) \nonumber \\ + P(\varepsilon_{\rm vib},2,kT)\sigma_1(E) .
\end{eqnarray}
Using the identity
\begin{eqnarray}
\sum_{i=0}^{\infty} P(\varepsilon_i,j_i,kT) \equiv [P(0,0,kT)+\sum_{i=2}^{\infty} P(\varepsilon_i,j_i,kT)] \nonumber \\ + P(\varepsilon_{\rm vib},2,kT) \equiv 1 \;
\end{eqnarray}
we reduce Eq. (26) to the simple form
\begin{eqnarray}
\sigma(E,kT) &\approx& 
\sigma_0(E)+P(\varepsilon_{\rm vib},2,kT)[\sigma_1(E)-\sigma_0(E)] \nonumber \\
&=& \sigma_0(E)\{1 + P(\varepsilon_{\rm vib},2,kT)[s(E)-1]\} . 
\end{eqnarray}
Here term containing $P(\varepsilon_{\rm vib},2,kT)$ is related to the cross-section enhancement induced by the population of the first $2^+$ surface oscillation state in soft nuclei in the stellar matter, and 
\begin{eqnarray}
s(E)=\sigma_1(E)/\sigma_0(E).
\end{eqnarray}
Ratio $s(E)$ directly shows the effect of cross-section enhancement caused by deformation of the nuclear surface in $2^+$ states, because if the surface deformation is neglected, then $\sigma_1(E)=\sigma_0(E)$ and $s(E)=1$. 

If the nuclear surface is spherical in the ground and excited states, then $\sigma_i(E)=\sigma_0(E)$ for any $i$ and $s(E)=1$. In this case we obtain from Eq. (3) using the identity (27) that
\begin{eqnarray}
\sigma_{\rm sph}(E,kT) = \sigma_0(E) .
\end{eqnarray}
This result we can also obtain from Eq. (28) in the limit $s(E)=1$. 

The reaction $S$-factor is proportional to the cross-section \cite{ns1}
\begin{eqnarray}
S(E,kT) =	E \exp(-2 \pi \eta(E)) \sigma(E,kT),
\end{eqnarray}
where $\eta(E)=z Z e^2/(\hbar v)$ is the Sommerfeld parameter, $v=(2E/\mu)^{1/2}$ is the relative velocity in the entrance channel. 

The enhancement of the $S$-factor or the reaction cross-section in stellar matter induced by $2^+$ surface oscillation is described by the ratio
\begin{eqnarray}
s(E,kT)=\frac{S(E,kT)}{S_{\rm sph}(E)} = \frac{ \sigma(E,kT)}{\sigma_{\rm sph}(E,kT)} \nonumber \\
\approx 1+ P(\varepsilon_{\rm vib},2,kT)[s(E)-1], 
\end{eqnarray}
where
\begin{eqnarray}
S_{\rm sph}(E) =	E \exp(-2 \pi \eta(E)) \sigma_{\rm sph}(E,kT) \nonumber \\ = E \exp(-2 \pi \eta(E)) \sigma_0(E). 
\end{eqnarray}

\begin{figure}
\includegraphics[width=9.0cm]{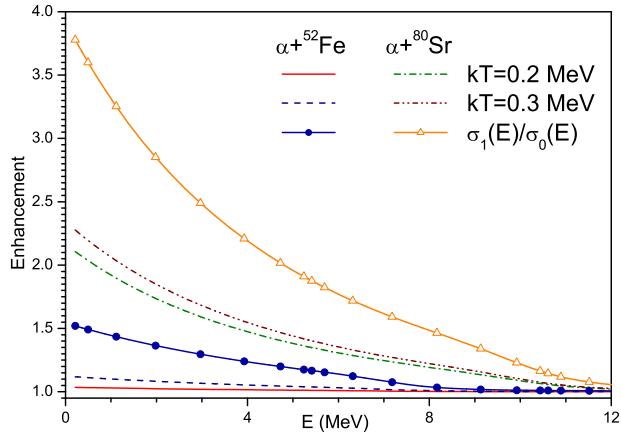}
\caption{(Color online) The $S$-factor enhancement for capture reactions $\alpha+^{52}$Fe and $\alpha+^{80}$Sr.}
\end{figure}

The results of $S$-factor enhancement for capture reactions $\alpha+^{52}$Fe and $\alpha+^{80}$Sr evaluated for star temperature $kT=0.2$ and 0.3 MeV are presented in Fig. 2. These temperatures are related to the O-Si burning phases in the core of massive stars \cite{rr,whw}. As seen in this figure the $S$-factor is enhanced by the $2^+$ surface oscillations at low collision energies $E$. The $S$-factor enhancement for reactions $\alpha+^{52}$Fe and $\alpha+^{80}$Sr grows with the temperature of the stellar matter in Fig. 2, because the partition of $2^+$ state increases in this temperature interval, see Fig. 1. 

We also present results for the cross-section ratio
$s(E)$ in Fig. 2.  In Fig. 2 we see that values of $s(E)$ are significantly larger then 1 for both reactions at low (sub-barrier) collision energies $E$. The cross-section enhancement induced by deformation is more important for $\alpha+^{80}$Sr system.

The dependence of the transmission coefficient on the surface deformation is negligible at collision energies above the barrier. Therefore $s(E,kT)$ and $s(E)$ are close to 1 at high collision energies, see Fig. 2. 

The stellar reaction cross-sections are often averaged over the Maxwell--Boltzmann distribution of collision velocities $v$ \cite{rr,s,r}. This cross-section can be presented in the form \cite{s}
\begin{eqnarray}
\langle \sigma(kT) \rangle = \frac{2}{\sqrt{\pi}} \frac{\int_0^\infty \sigma(E,kT)E \exp(-E/kT) dE}{\int_0^\infty E \exp(-E/kT) dE} ,
\end{eqnarray}
where brackets $\langle \rangle$ mean the Maxwell--Boltzmann averaging over the collision velocities. We determine the velocity-averaged cross-sections $\langle \sigma_{\rm sph}(kT) \rangle$, $\langle \sigma_0(kT) \rangle$ and $\langle \sigma_1(kT) \rangle$ in the similar way. Note that $\langle \sigma_{\rm sph}(kT) \rangle=\langle \sigma_0(kT) \rangle$ due to Eq. (30).

\begin{figure}
\includegraphics[width=9.0cm]{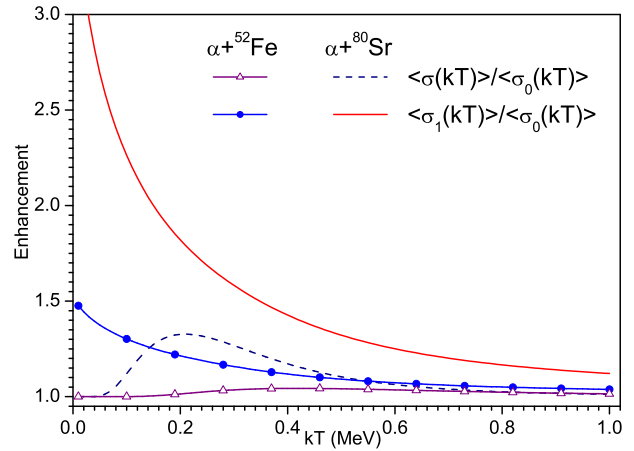}
\caption{(Color online) The enhancement of velocity-averaged cross-sections of $\alpha$-capture reactions $\alpha+^{52}$Fe and $\alpha+^{80}$Sr.}
\end{figure}

The enhancement of velocity-averaged cross-sections $\langle \sigma(kT) \rangle/\langle \sigma_{\rm sph}(kT) \rangle=\langle \sigma(kT) \rangle/\langle \sigma_0(kT) \rangle$ for $\alpha$-capture reactions $\alpha+^{52}$Fe and $\alpha+^{80}$Sr caused by $2^+$ shape vibrations is shown in Fig. 3. Ratios of velocity-averaged $\alpha$-capture cross-sections by nucleus with vibrating and spherical surfaces $\langle \sigma_1(kT) \rangle/\langle \sigma_0(kT) \rangle$ are also presented in this figure for comparison. As seen in Fig. 3 the surface oscillations enhance the velocity-averaged cross-sections. Due to this the ratio $\langle \sigma(kT) \rangle/\langle \sigma_0(kT) \rangle$ is noticeably larger 1 for $kT \gtrsim 0.1$ MeV. However for low temperature of star matter $kT \lesssim 0.1$ MeV the partition probability of $2^+$ surface oscillation state is very small (see in Fig. 1), therefore $\langle \sigma(kT) \rangle/\langle \sigma_0(kT) \rangle$ is close to 1. At star matter temperatures $kT \gtrsim 0.6$ MeV the main contribution to the velocity-averaged capture cross-section is related to higher collision energies $E$, which is close to the barrier. Therefore the effect of the barrier penetration enhancement due to surface vibrations of nucleus in $2^+$ state is diminished. 

When the results presented in Figs. 2 and 3 are compared, it is apparent that the $2^+$ surface vibrations are important for both the $S$-factor values at low energies and the velocity-averaged cross-sections at moderate star temperatures. The enhancement of the reaction $S$-factor or the capture cross-section induced by $2^+$ nuclear surface excitation in stellar matter can change the balance between various reactions. This may affect the equilibrium conditions between direct and inverse reactions as well as the abundance of elements in hot stellar matter. 

\section{$p$-capture reactions in stars}

We consider $p$-capture reactions on either a spherical or deformed nucleus in hot star matter using similar formalism as that for the $\alpha$-capture reactions in the previous section. The spin of proton is $j_p= {1 \over 2}$. As a result the nuclear part of the $p$-nucleus potential $v_n(r,\theta)$ should include the spin-orbit contribution \cite{chepurnov,nemirovski} in contrast to the $\alpha$-particle case. Therefore we should modify the approach for cross-section evaluation presented in the previous section: 
\begin{itemize}
\item[1.] The nuclear part of $\alpha$-nucleus potential described by Eqs. (16)-(22) is substituted by the Chepurnov $p$-nucleus potential with the central and spin-orbit parts \cite{chepurnov}. The central part of the Chepurnov $p$-nucleus potential has also the Woods-Saxon shape (16), while its spin-orbit part is proportional to the radial derivative of the Woods-Saxon potential. 

\item[2.] The deformation of nuclear surface is taken into account in the spin-orbit potential too. 

\item[3.] Summation over $J$ in Eq. (4)
\begin{eqnarray}
\frac{1}{2j_p+1}\sum_{J \ell \ell'} (2J+1) t_{J \ell \ell'}(E,\ell,\theta) \\ = \frac{1}{2} \sum_{J \ell \ell'} (2J+1) t_{J \ell \ell'}(E,\ell,\theta) \nonumber
\end{eqnarray}
is replaced by (see \cite{satchler,nemirovski})
\begin{eqnarray}
\sum_\ell \left[(\ell+1) \; t_{J=\ell+\frac{1}{2}}(E,\ell,\theta)+\ell \; t_{J=\ell-\frac{1}{2}}(E,\ell,\theta)\right] ,
\end{eqnarray}
where $t_{J}(E,\ell,\theta)$ is the transmission coefficient, which shows the probability of proton penetration through the barrier formed at the angle $\theta$ between the symmetry axis of axial-symmetric deformed nuclei and the vector directed from the center of the deformed nucleus to the proton, and $J$ is the total angular momentum ${\vec J}={\vec \ell}+\vec{j_p}$.
\end{itemize}

\begin{figure}
\includegraphics[width=9.0cm]{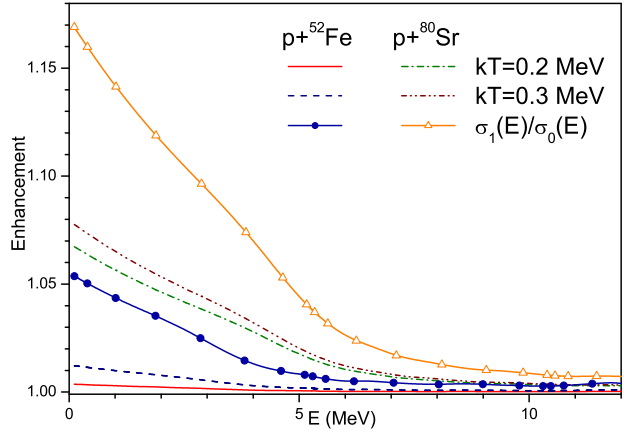}
\caption{(Color online) The $S$-factor enhancement for capture reactions $p+^{52}$Fe and $p+^{80}$Sr. }
\end{figure}

Note that Eq. (36) is exact in the case of $p$-capture on a nucleus in the state with zero value of the spin (see Appendix A in Ref. \cite{satchler}). We would like to estimate the deformation affect $p$-capture roughly. Therefore, for the sake of simplicity, we neglect the spin of nuclear states in a heavy nucleus and apply Eq. (36) for $p$-capture on the nucleus with any value of spin. 

From Fig. 4, we can see that the $2^+$ surface oscillations weakly affect the $S$-factor values for reactions $p+^{52}$Fe and $p+^{80}$Sr. The enhancement of $p$-capture cross-sections averaged over the Maxwell--Boltzmann distribution of collision velocities induced by surface vibrations for these reactions are given in Fig. 5. As seen the $2^+$ surface oscillations slightly enhance the velocity-averaged cross-sections for $p$-capture reactions. 

\begin{figure}
\includegraphics[width=9.0cm]{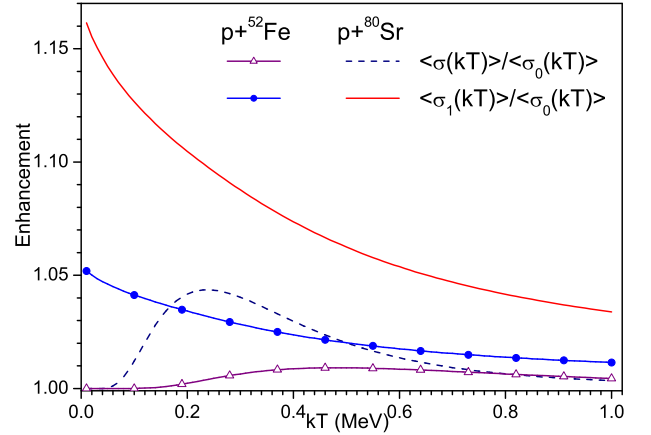}
\caption{(Color online) The enhancement of velocity-averaged cross-sections of $p$-capture reactions $p+^{52}$Fe and $p+^{80}$Sr.}
\end{figure}

\section{Nucleus-nucleus fusion reactions in stars}

Both nuclei participating in fusion reactions can be excited in hot stellar matter. Therefore the nucleus-nucleus fusion cross-sections can be evaluated as
\begin{eqnarray}
\sigma(E,kT)= \sum_{i_1,i_2=0}^{\infty} P_1(\varepsilon_{i_1},j_{i_1},kT) P_2(\varepsilon_{i_2},j_{i_2},kT) \nonumber \\ \times \sigma_{i_1 i_2}(E),
\end{eqnarray}
where $P_1(\varepsilon_{i_1},j_{i_1},kT)$ and $P_2(\varepsilon_{i_2},j_{i_2},kT)$ are the partition probabilities of states $i_1$ in the first nucleus and $i_2$ in the second one, respectively, $\sigma_{i_1 i_2}(E)$ is the fusion reaction cross-section between these nuclei in the states $i_1$ and $i_2$.

If one nucleus is rigid, while another one is soft, the $2^+$ surface oscillations are mainly important in the soft reaction partner. In such case the double sum in Eq. (37) reduces to the single sum, see Eq. (2). Therefore we can easy adapt our approach for the cross-section evaluation presented in previous sections to the nucleus-nucleus fusion reactions. 

Let's consider fusion reaction $^{16}$O+$^{32}$Mg. This reaction may take place in massive stars during explosive O-Si burning phases. The neutron-reach $^{32}$Mg can be formed in hot star matter by $n$-capture. Compound-nucleus formed in this reaction is $^{48}$Ca.

Nucleus $^{16}$O is spherical, double-magic and very rigid. The energies of the lowest $0^+$, $3^-$ and $2^+$ states in $^{16}$O are, respectively, 6.05, 6.13 and 6.92 MeV \cite{be2,nudat}. Therefore probabilities of population of such high-energy states in stars at $kT \lesssim 1$ MeV are negligible. The ground-state shape of $^{32}$Mg is spherical \cite{msmn}. The excitation energy of the lowest $2^+$ surface oscillation state in $^{32}$Mg is $\varepsilon_{\rm vib}=0.8855$ MeV \cite{be2}. The total vibrational amplitude of this state is very large $\beta_{\rm vib}=0.473$ \cite{be2}. (Data for other exited states are picked up from Ref. \cite{nudat}.) So, the neutron-reach nucleus $^{32}$Mg is very soft. In this case the fusion reaction model described in Sec. III may be applied with minor modifications to reaction $^{16}$O+$^{32}$Mg. The modifications are only related to the nuclear part of the interaction potential. 

The functional form of the nuclear interaction potential between spherical nuclei has been discussed in Ref. \cite{d}. This potential is obtained by using the semi-microscopic calculations of the interaction energy of two nuclei. Heights and radii of empirical fusion barriers are well described by this potential for various pairs of interacting nuclei \cite{d,dt,newton}. 

\begin{figure}
\includegraphics[width=9.0cm]{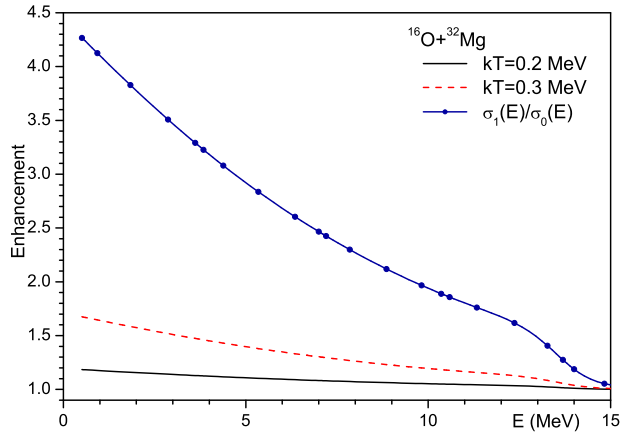} 
\caption{(Color online) The $S$-factor enhancement for fusion reactions $^{16}$O+$^{32}$Mg.}
\end{figure}

The nuclear part of the potential depends on both the distance between surfaces of colliding nuclei $d(r,\theta)$ and the surface curvature of corresponding nuclei at the closest point \cite{prox}. Therefore the nuclear part of the potential between spherical and deformed nuclei may be approximated as
\begin{eqnarray}
v_n(d(r,\theta)) \approx 
(C_1+C_2)/(C_1+C_2(\theta)) \; v_n^{0}(d(r,\theta)).
\end{eqnarray}
Here $v_n^{0}(d(r))$ is the nuclear part of the interaction potential between two spherical nuclei with radii $R_1^0$ and $R_2^0$ at distance between their surfaces $d(r)=r-R_1^0-R_2^0$ \cite{d}, $C_{1(2)}=1/R_{1(2)}^0$ are the surface curvatures of corresponding spherical nuclei, $R_2(\theta)=R_2^0 (1+ \beta Y_{20}(\theta))$ is the distance of the surface from the origin of deformed nucleus, $C_2(\theta) \approx (1+2 \beta Y_{20}(\theta))/R_2^0$ is the curvature of deformed nuclear surface, and $d(r,\theta) \approx r-R_1^0-R_2(\theta)$. Substituting potential (38) into (13) we can estimate the cross-section of heavy-ion fusion reaction in the framework of our model. 

The $S$-factor enhancement for fusion reaction $^{16}$O+$^{32}$Mg induced by the $2^+$ surface oscillation in $^{32}$Mg is shown in Fig. 6. The $S$-factor values are strongly enhanced by the surface oscillations especially at low collision energies $E$. The enhancement of the fusion cross-section for this reaction averaged over the Maxwell--Boltzmann distribution of collision velocities is presented in Fig. 7. As seen in Fig. 7 the $2^+$ surface oscillations increase the velocity-averaged cross-sections of nucleus-nucleus fusion reactions in stellar matter. 

\begin{figure}
\includegraphics[width=9.0cm]{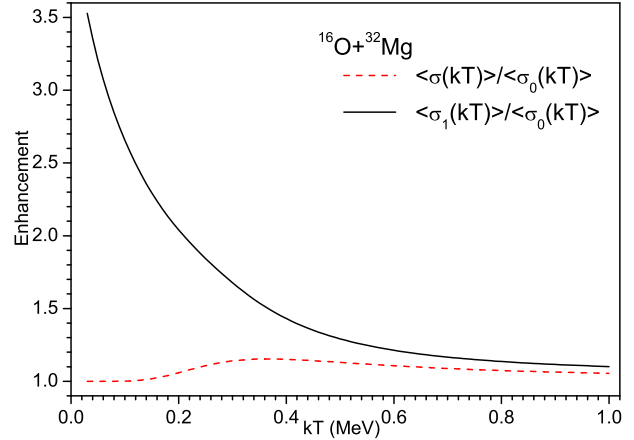}
\caption{(Color online) The enhancement of velocity-averaged cross-sections for reaction $^{16}$O+$^{32}$Mg.}
\end{figure}

\section{Capture of charged particles on nuclei with well-deformed ground-state in stars}

If a nucleus is well-deformed in the ground-state, then two kinds of surface deformations (static one of the ground-state and dynamic one of the $2^+$ vibrational state) should be taken into account in evaluation of the interaction potential between a charged particle and the nucleus. The dynamical deformations are related to quadrupole surface vibrations about the deformed ground-state shape of the nucleus. Both types of quadrupole surface deformations should affect the cross-sections of nuclear reactions in hot stellar matter. 

We can extend our model for evaluation the capture reactions to a nucleus with deformed ground-state shape. The extension to such case is straightforward. The statical amplitude $\beta_{\rm statical}$ should be taken into account for evaluation of capture on the nucleus in the ground state $\sigma_0(E)$, while the statical and dynamical deformation amplitudes should be summed up for estimation of capture on the nucleus in the first $2^+$ vibrational state $\sigma_1(E)$. By putting $D(\beta)=\delta(\beta-\beta_{\rm statical})$ we adapt Eqs. (7), (10) and (12) for obtaining $\sigma_0(E)$, see also Ref. \cite{di}. The influence of the $2^+$ surface oscillations around the deformed ground-state shape on physical quantities will be similar to the case of the $2^+$ surface vibrations around spherical shape. Thus we should substitute $D(\beta)$ in Eq. (10) by $D(\beta-\beta_{\rm statical})$ for estimating the effect of the $2^+$ surface oscillations around the deformed shape. Here we take into account that vibrations of $\beta$-type about deformed shape \cite{soloviev} give the largest affect the particle-nucleus potential and that the nature of quadrupole surface oscillations in deformed and spherical nuclei is similar.

\subsection{$\alpha$-capture}

Reactions $\alpha$+$^{22}$Ne and $\alpha$+$^{24}$Mg are very important for the burning of massive star and nucleosynthesis in stellar matter \cite{whw,ra}.  

\begin{figure}
\includegraphics[width=9.0cm]{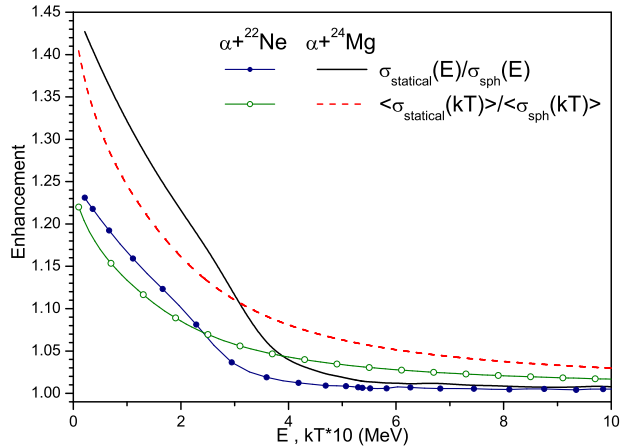}
\caption{(Color online) The enhancement of $S$-factor and velocity-averaged cross-section for reactions $\alpha$+$^{22}$Ne and $\alpha$+$^{24}$Mg induced by statical deformation of heavy nuclei.}
\end{figure}

Nuclei $^{22}$Na and $^{24}$Mg are well-deformed in the ground-state. The values of the corresponding deformation parameter are $\beta_{\rm statical}=0.326$ and $0.374$ \cite{msmn}. Excitation energies of the first $2^+$ state in these nuclei are respectively 1.27 and 1.37 MeV \cite{be2}. However the nature of these excited states is rotational. Energies of the lowest $2^+$ vibrational state are correspondingly 4.46 and 4.24 MeV \cite{nudat}. Energies are sufficiently high, therefore the occupation probabilities of the lowest $2^+$ vibrational excitation in $^{22}$Na or $^{24}$Mg in stellar matter at temperatures $kT \lesssim 0.3$ MeV are small. Consequently, we can ignore the dynamical deformations related to the $2^+$ vibrational state in these nuclei and take into account only the statical deformation. Shapes of the nucleus in the ground and low-spin rotational excited states are very similar, therefore the cross-sections for the these states are the same. Occupation probabilities of the ground and low-spin rotational excited states are summed and the net probability of these states is very close to 1.

We present the evaluation of ratio $\sigma_{\rm statical}(E)/\sigma_{\rm sph}(E)$ for reactions $\alpha$+$^{22}$Ne and $\alpha$+$^{24}$Mg in Fig. 8, where $\sigma_{\rm statical}(E)$ and $\sigma_{\rm sph}(E)$ are the cross-sections obtained respectively for deformed and spherical shapes of heavy nuclei. We see that the enhancement of the cross-sections or the $S$-factors and the velocity-averaged cross-sections for reactions $\alpha$+$^{22}$Ne and $\alpha$+$^{24}$Mg induced by statical deformation is strong. This enhancement increases as the temperature of stellar matter decreases.

Comparing results in Fig. 8 we conclude that enhancement of capture cross-section caused by statical deformation of Mg is lager then that for Ne case. There are two reasons for such effect. Firstly, the value of the deformation parameter in $^{24}$Mg is larger then that in $^{22}$Ne. Secondly, the  number of protons in $^{24}$Mg is also larger then that in $^{22}$Ne. 

The range of barrier distribution (24) or (25) for $\alpha$-capture reaction is induced by both the statical deformation of heavy nucleus $\beta$ and the mutual orientation of the heavy nucleus and the incoming $\alpha$-particle (angle $\theta$). In contrast to this the treatment of $\alpha$-capture reactions on deformed nuclei based on an effective spherical potential of equal volume is related to a single barrier. The height of this single barrier $v_{\rm bar}^{\rm sph}(\beta)$ estimated at deformation $\beta$ is slightly smaller then that evaluated for the same colliding system proceeding from systematics for the $\alpha$-nucleus potential $v_{\rm bar}^{\rm sph. \; syst.}$. The barrier reduction due to volume correction is proportional to $\beta^2$ for effective spherical potential of equal volume, while variation of the barrier induced by $\beta$ in our approach (see Eqs. (14)-(22)) is proportional to $\beta$ and depends on $\theta$. As a result the values of potentials obey inequalities
\begin{eqnarray*}
 v_{\rm bar}(\theta=0)|_{\beta>0} < v_{\rm bar}^{\rm sph}(\beta) < v_{\rm bar}^{\rm sph. \; syst.} < v_{\rm bar}(\theta=\frac{\pi}{2})|_{\beta>0} .
\end{eqnarray*}
Therefore the transmission coefficients $t(E,\ell,\theta)$ and the ratio $\sigma_{\rm statical}(E)/\sigma_{\rm sph}(E)$ are differently described in the frameworks of these two approaches at very low collision energies $E$. The difference between cross-section values evaluated in both approaches rises with the value of deformation parameter.

\subsection{$p$-capture}

\begin{figure}
\includegraphics[width=9.0cm]{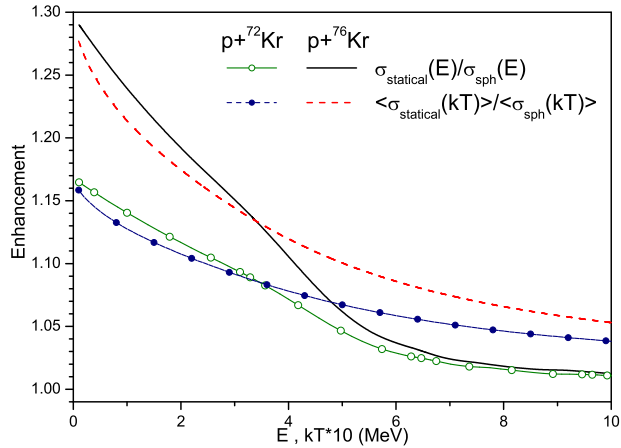}
\caption{(Color online) The enhancement of $S$-factor and velocity-averaged cross-section for reaction $p$+$^{72,76}$Kr induced by statical deformation of heavy nucleus.}
\end{figure}

The $p$-capture reactions on proton-rich nuclei are important for $rp$-nucleosynthesis in stellar matter \cite{rr,rp}. Therefore let us consider the reactions $p$+$^{72,76}$Kr. The $^{72}$Kr is a significant waiting point in the $rp$-process \cite{rp}. Proton-rich nuclei $^{72,76}$Kr are well-deformed in the ground-state. It is interesting that the ground-state shape is oblate for $^{72}$Kr and prolate for $^{76}$Kr. Values of the corresponding deformation parameter are $\beta_{\rm statical}=-0.349$ and 0.4 \cite{msmn}. 
  
The $p$-capture reaction below barrier is also strongly enhanced by the ground-state deformation of nuclei, see Fig. 9. We neglect the dynamical deformation related to the $2^+$ vibrational state in $^{72,76}$Kr, because energies of the lowest vibrational states in these nuclei are much larger then the energies of the lowest rotational levels \cite{nudat}. Due to this, vibrational levels in these nuclei are weakly populated in stellar matter. The spin of proton is exactly taken into account in our evaluation in this case, because the ground-state spin of even-even nuclei $^{72,76}$Kr is zero and we neglect the dynamical deformations. Comparing results in Fig. 9 we see that the prolate deformation stronger enhances the $p$-capture reaction cross-section than similar oblate deformation. 

\subsection{Heavy-ion fusion}

The enhancement of $S$-factor and velocity-averaged cross-section for reaction $^{16}$O+$^{22}$Ne induced by statical deformation of heavy nucleus is presented in Fig. 10.
The properties of excited states in the colliding nuclei are discussed before. Therefore we take into account only statical ground-state deformation of $^{22}$Ne in our calculations. Reaction $^{16}$O+$^{22}$Ne may take place at O-Si burning phase in hot stellar matter. We see that the enhancement of both the $S$-factor and the velocity-averaged cross-section is very strong especially at low temperatures of stellar matter. 

\begin{figure}
\includegraphics[width=9.0cm]{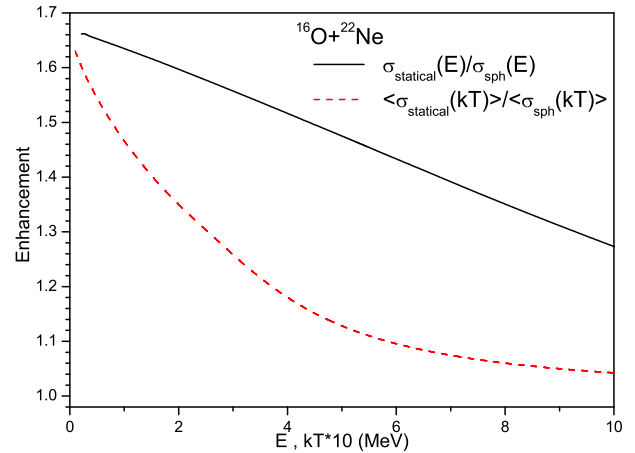}
\caption{(Color online) The enhancement of $S$-factor and velocity-averaged cross-section for reaction $^{16}$O+$^{22}$Ne induced by statical deformation of heavy nucleus.}
\end{figure}

\section{Discussion and conclusions}

We demonstrated that the fusion (capture) reactions between two nuclei or between a charged particle and a nucleus at sub-barrier collision energies are enhanced by the $2^+$ surface oscillations in soft nuclei in hot stellar matter. The cross-sections averaged over collision velocities are also increased by the $2^+$ surface vibrations. The effect of cross-section enhancement, induced by the shape oscillations, depends on both the temperature of stellar matter and the excitation energy of $2^+$ state. 

Comparing the results for $\alpha$-, $p$-capture reactions and heavy-ion fusion reactions we see that the influence of the $2^+$ surface vibrations grows with increasing charges and masses of the colliding particles. This influence is negligible for the $p$-capture and important for the $\alpha$-capture and heavy-ion fusion reactions.

The enhancement of capture cross-section by the quadrupole surface oscillations is significant for soft nuclei with low-energy $2^+$ states. Such nuclei are often occurred far from the $\beta$-stability line. This enhancement of capture rate is negligible in the case of rigid nuclei. Therefore the mass and charge dependencies of the lowest $2^+$ state are significant for abundances of heavy proton-rich elements in hot star matter \cite{rp,ra}. Reactions with heavier colliding partners are stronger enhanced by the low-energy $2^+$ oscillations, because the barrier reduction induced by axial quadrupole oscillations of nuclear surface are larger in heavier systems.

Our model can be easy extended to the case of high-multipole $\lambda \geq 3$ axial surface oscillations. However the energies of high-multipole surface vibrations are higher than those for $\lambda=2$ in the most part of nuclei \cite{be2}. Moreover the amplitude of the potential barrier reduction due to axial $\lambda \geq 3$ or nonaxial $\lambda \geq 2$ multipole surface distortions are smaller than that caused by the axial quadrupole deformation. So the effect of high-multipole axial or nonaxial nuclear surface vibrations may be significant only for the charged particle capture on nuclei, which are especially soft to a specific kind of surface deformation.  

We have studied the enhancement of fusion reaction cross-section induced by the $2^+$ surface oscillations of nuclei participating in reactions in the star matter. However, many other mechanisms of sub-barrier fusion enhancement have been discussed, see Refs. \cite{oo,fl,wong,dr,newton,subfus} and references therein. Therefore it is interesting to evaluate the role of various reaction mechanisms in the framework of detailed models. 

The enhancement of reaction cross-sections, induced by the surface oscillations of colliding nuclei, is important for careful evaluation of the equilibrium composition of nuclei in hot stellar matter. The $2^+$ surface oscillations in soft nuclei with spherical ground-state shape should be taken into consideration of the nucleosynthesis during supernova explosions. 

It is shown that the $p$-, $\alpha$-capture on nuclei with well-deformed ground-state are significantly enhanced by the ground-state deformation of nuclei at low temperatures of stellar matter. Similar effect also takes place for fusion reactions between deformed heavy-ions. This effect is very important for the O-Si burning phases of massive stars, when reactions between charged particles and well-deformed nuclei such as $^{19-21}$F, $^{20-24}$Ne, $^{20-26}$Ne, $^{22-28}$Mg, $^{22-29}$Al and $^{24-29}$Si determine burning processes in massive stars. Enhancement of the charged particle capture reactions by the ground-state deformation is very significant for the nucleosynthesis of proton-rich elements in hot stellar matter. The origin of the large abundances of proton-rich nuclei with well-deformed shapes, especially in the Z = 36-42 region \cite{rp}, may be associated with enhancement of the charge-particle capture cross-section caused by the ground-state deformation.

Comparing results for the velocity-averaged cross-sections in Figs. 3, 5, 7 and 8-10 we see that the enhancement of the reaction cross-section caused by statical deformation is very strong for low temperatures of star matter. However the deformation effect for reactions with soft spherical nuclei is important for high temperatures, when the occupation probability of the lowest $2^+$ vibrational state is significant. The enhancement of the reaction cross section related to the statical deformation is very strong at low temperature, because the ground-state is mainly populated at low temperatures.

The author thanks P. Danilewicz, A. Ya. Dzyublik, Yu. B. Ivanov, A. G. Magner and Referee for useful remarks.


\begin{references}

\bibitem{st} S. L. Shapiro, S. A. Teukolsky, {\it Black holes, white dwarfs, and neutron stars} (John Wiley and Sons, New York, 1983).

\bibitem{rr} C. E. Rolfs, W. S. Rodney, {\it Cauldons in the Cosmos} (The university of Chicago Press, Chicago Press, 1988).

\bibitem{bethe} H. A. Bethe, Rev. Mod. Phys. {\bf 62}, 801 (1990).

\bibitem{lw} K. Langanke, M. Wiescher, Rep. Prog. Phys. {\bf 64}, 1657 (2001).

\bibitem{whw} S. E. Woosley, A. Heger, T. A. Weaver, Rev. Mod. Phys. {\bf 74}, 1015 (2002).

\bibitem{ns1} G. Wallerstein, I. Iben, P. Parker, A. M. Boesgaard, G. M. Hale, A. E. Champagne, Ch. A. Barnes, F. Kappeler, V. V. Smith, R. D. Hoffman, F. X. Timmes, Ch. Sneden, R. N. Boyd, B. S. Meyer, D. L. Lambert, Rev. Mod. Phys. {\bf 69}, 995 (1997). 

\bibitem{rp} H. Schatz, A. Aprahamian, J. G\"orres, M. Wiescher, T. Rauscher, J. F. Remges, F.-K. Thielemann, B. Pfeiffer, P. M\"oller, K.-L. Kratz, H. Herndl, B. A. Brown, H. Rebel, Phys. Rep. {\bf 294}, 167 (1998).

\bibitem{ra} T. Rauscher, F.-K. Thielemann, J. G\"orres, M. Wiescher, Nucl. Phys. A {\bf 675}, 695 (2000).

\bibitem{ns3} F.-K. Thielemann, F. Brachwitz, C. Freiburghaus, E. Kolbe, G. Martinez-Pinedo, T. Rauscher, F. Rembges, W. R. Hix, M. Liebendorfer, A. Mezzacappa, K.-L. Kratz, B. Pfeiffer, K. Langanke, K. Nomoto, S. Rosswog, H. Schatz, W. Wiescher, Progr. Part. Nucl. Phys. {\bf 46}, 5 (2001).

\bibitem{ns2} Y.-Z. Qian, Progr. Part. Nucl. Phys. {\bf 50}, 153 (2003).

\bibitem{sr} G. R. Caughlan, W. A. Fowler, At. Data Nucl. Data Tabl. {\bf 40}, 283 (1988).

\bibitem{s} Z. Y. Bao, H. Beer, F. Kapper, F. Voss, K. Wisshak, T. Rauscher, At. Data Nucl. Data Tabl. {\bf 76}, 70 (2000).

\bibitem{r} F.-K. Thielemann, T. Rauscher, At. Data Nucl. Data Tabl. {\bf 75}, 1 (2000); {\bf 79}, 47 (2001).

\bibitem{oo} P.-G. Reinhard, J. Friedrich, K. Goeke, F. Gr\"ummer, D. H. E. Gross, Phys. Rev. C {\bf 30}, 878 (1984).

\bibitem{fragm} A. S. Botvina, I. N. Mishustin, Phys. Rev. C {\bf 72}, 048801 (2005).

\bibitem{beta} K. Langanke, G. Martinez-Pinedo, At. Data Nucl. Data Tabl. {\bf 79}, 1 (2001).

\bibitem{be2} S. Raman, C. W. Nestor, P. Tikkanen, At. Data Nucl. Data Tabl. {\bf 78}, 1 (2001).

\bibitem{satchler} G. R. Satchler, {\it Direct nuclear reactions} (Claredon press, Oxford, 1983).

\bibitem{fl} P. Fr\"obrich, R. Lipperheide, {\it Theory of nuclear reactions} (Claredon press, Oxford, 1996).

\bibitem{dn} V. Yu. Denisov, W. N\"orenberg, Eur. Phys. J. {\bf A15}, 375 (2002).

\bibitem{wong} C. Y. Wong, Phys. Rev. Lett. {\bf 31}, 766 (1973).

\bibitem{dr} V. Yu. Denisov, G. Royer, J. Phys. G {\bf 20}, L43 (1994); V. Yu. Denisov, S.V. Reshitko, Phys. At. Nucl., {\bf 59}, 78 (1996).

\bibitem{di} V. Yu. Denisov, H. Ikezoe, Phys. Rev. C {\bf 72}, 064613 (2005).

\bibitem{ccdef} J. O. Fernandez-Niello, C. H. Dasso, S. Landowne, Comp. Phys. Commun. {\bf 54}, 409 (1989).

\bibitem{d} V. Yu. Denisov, Phys. Lett. {\bf B526}, 315
(2002).

\bibitem{nudat} http://www.nndc.bnl.gov/nudat2/.

\bibitem{soloviev} V. G. Soloviev {\it The theory of atomic nucleus: Nuclear models} (Energoizdat, Oxford, 1981).

\bibitem{subfus} 
H. Esbensen, Nucl. Phys. A {\bf 352}, 147 (1981); 
M. Beckerman, Rep. Prog. Phys. {\bf 51}, 1047 (1988); M. Dasgupta, D. J. Hinde, N. Rowley, A. M. Stefanini, Annu. Rev. Nucl. Part. Sci. {\bf 48}, 401 (1998); A. B. Balantekin, 
N. Takigawa, Rev. Mod. Phys., {\bf 70}, 77 (1998); V. Yu. Denisov, Phys. At. Nucl. {\bf 62}, 1349 (1999); V. Yu. Denisov, Eur. Phys. J. A {\bf 7}, 87 (2000); L. F. Canto, P. R. S. Gomes, R. Donangelo, M. S. Hussein, Phys. Rep. {\bf 424}, 1 (2006). 

\bibitem{chepurnov} V. A. Chepurnov, Yad. Fiz. {\bf 6}, 955 (1967).

\bibitem{nemirovski} P. E. Nemirovski, {\it Modern models of atomic nucleus}, (Atomizdat, Moscow, 1960).

\bibitem{msmn} P. M\"oller, J. R. Nix, W. D. Myers, and W. J. Swiatecki, At. Data Nucl. Data Tables {\bf 59}, 185 (1995).

\bibitem{newton} J. O. Newton, R. D. Butt, M. Dasgupta, D. J. Hinde, I. I. Gontchar, C. R. Morton, K. Hagino, Phys. Rev. C {\bf 70}, 024605 (2004).

\bibitem{dt} V. Yu. Denisov, AIP Conference Proc. {\bf 704}, 92 (2004); arXiv:nucl-th/0310019.

\bibitem{prox} J. Blocki, J. Randrup, W.J. Swiatecki,
C.F. Tsang, Ann. Phys. {\bf 105}, 427 (1977).

\end{references}
\end{document}